\input phyzzx

\def\dplus{=\hskip-5pt \raise 0.7pt\hbox{${}_\vert$} ^{\phantom 7}}
\def\dplusup{=\hskip-5.1pt \raise 5.4pt\hbox{${}_\vert$} ^{\phantom 7}}
\def\dplus{=\hskip-4.8pt \raise 0.7pt\hbox{${}_\vert$} ^{\phantom 7}}

\def\pmb#1{\setbox0=\hbox{#1} \kern-.025em\copy0\kern-\wd0
\kern0.05em\copy0\kern-\wd0 \kern-.025em\raise.0433em\box0}

\def\th{\theta}\def\i{\iota}

\REF\aat{A.A. Tseytlin, Class. Quantum. Grav. {\bf 12} (1995) 2365.}
\REF\cmh{S.J. Gates, C.M. Hull and M. Ro\v cek, Nucl. Phys. {\bf B248}
(1984) 157.}
\REF\hpa {P.S. Howe and G. Papadopoulos, Nucl .Phys. {\bf B289} (1987) 264;
Class. Quantum Grav. {\bf 5} (1988) 1647; Nucl .Phys. {\bf B381} (1992) 360.}
\REF\sal{S. Salamon, Invent Math. {\bf 67} (1982), 143; Ann. Sci. ENS Supp. {\bf 19} (1986).}
\REF\roc{N.J. Hitchin, A. Karlhede, U. Lindstr\"om, and M. Ro\v cek, Commun. Math. Phys. {\bf 108}
(1987) 535.}
\REF\dks{F. Delduc, S. Kalitsin and E. Sokatchev, Class. Quantum. Grav. {\bf 7}
(1990) 1567.}
\REF\hw{C.M. Hull and E. Witten, Phys. Lett. {\bf 160B} (1985) 398.}
\REF\sen{A. Sen, Nucl. Phys. {\bf B278} (1986) 289. }
\REF\hpc{ P.S. Howe and G. Papadopoulos, Class. Quantum Grav.
{\bf 4} (1987) 1749.}
\REF\vp {Ph. Spindel, A. Sevrin, W. Troost and A. Van Proeyen, Nucl. Phys. {\bf B308} (1988) 662;
{\bf B311} (1988) 465.}
\REF\iv{E.A. Ivanov, S.O. Krivonos and V.M. Leviant, Nucl. Phys. {\bf B304} (1988) 601; E.A. Ivanov
and S.O. Krivonos, J. Phys. A:{\bf 17} (1984) L671.} 
\REF\chs{C.G. Callan, J.A. Harvey and A. Strominger, Nucl. Phys. {\bf B359} (1991) 611.}
\REF\gp{G. Papadopoulos, Phys. Lett. {\bf B356} (1995) 249. } 
\REF\dv{F. Delduc and G. Valent, Class. Quantum Grav. {\bf 10} (1993) 1201.}
\REF\bv{G. Bonneau and G. Valent, {\sl Local heterotic geometry in holomorphic co-ordinates},
hep-th/9401003.}
\REF\gh {G.W. Gibbons and S.W. Hawking, Phys. Lett.  {\bf 78B} (1978) 430.}
\REF\gwg{G.W. Gibbons and P. J. Ruback, Commun. Math. Phys. {\bf 115} (1988) 267.}
\REF\hpb{P.S. Howe and G. Papadopoulos, Commun. Math. Phys. {\bf 151}
(1993) 467.}
\REF\rs{A.A. Rossly and A.S. Schwarz, Commun.Math.Phys. {\bf 105} (1986) 645.}
\REF\hh{P.S. Howe and G.G. Hartwell, Class. Quantum Grav. {\bf 12} (1995) 1823.}
\Pubnum{ \vbox{ \hbox{R/96/4}\hbox{CERN-TH/96-46} } }
\pubtype{}
\date{February, 1996}
\titlepage
\title{Twistor spaces for HKT manifolds}
\author{P.S. Howe\foot{Permanent address: Dept. of Mathematics, King's College London, London
WC2R 2LS} }
\address{CERN,  Theory Division \break CH-1211, Geneva \break Switzerland}
\andauthor{G. Papadopoulos}
\address{D.A.M.T.P \break 
 Silver Street \break Cambridge CB3 9EW}
\abstract {We construct the twistor space associated with an HKT
manifold, that is, a hyper-K\"ahler manifold with torsion, a type of
geometry that arises as the target space geometry in two-dimensional
sigma models with (4,0) supersymmetry. We show that this twistor space
has a natural complex structure and is a  holomorphic fibre
bundle over the complex
projective line with fibre the associated HKT manifold. We also show  
how the metric and torsion of the HKT manifold can be
determined from data on the twistor space by a reconstruction theorem.
We give a geometric description of the sigma model (4,0) superfields
as holomorphic maps (suitably understood) from a twistorial extension
of (4,0) superspace (harmonic superspace) into the twistor space of
the sigma model target manifold and write an action for the sigma
model in terms of these (4,0) superfields.}

\vskip 0.3cm

\endpage
\pagenumber=2
\font\mybb=msbm10 at 12pt
\def\bb#1{\hbox{\mybb#1}}

\def\I{\bb{I}}
\def\Z{\bb{Z}}

\def\J{\rlap J\mkern3mu{\rm J}}
\def\C{\mkern1mu\raise2.2pt\hbox{$\scriptscriptstyle|$}\mkern-7mu{\rm C}}

\def\cp{\C P}

\def\om{\omega}
\def\i{\iota}

Two-dimensional (p,0)-supersymmetric sigma models with Wess-Zumino term (torsion) are used to
describe the propagation of superstrings in curved backgrounds and arise naturally in the context of
heterotic string compactifications (for a recent review see [\aat]). These models have as couplings the
metric, $g$, of the target space,
$M$, and a locally defined two form, $b$, on $M$. Extended supersymmetry ($p\geq2$) imposes
restrictions on the couplings $g$ and $b$ of the sigma model which have an interpretation as
conditions on the geometry of the sigma model manifold.  In the absence of torsion, the geometry of
the sigma model manifolds is K\"ahler  or hyper-K\"ahler  depending on the number of supersymmetries
that leave the sigma model action invariant.  In the presence of torsion, the geometry of the sigma
model manifolds is {\it not} K\"ahler or hyper-K\"ahler and new geometry arises [\cmh,\hpa].
These new geometries, which we shall call K\"ahler with torsion (KT) and hyper-K\"ahler with
torsion (HKT), are, however, closely related to K\"ahler and hyper-K\'ahler geometries
respectively.
In this letter, we show that both KT and HKT geometries can be characterised in terms of
the properties of two-forms just as K\"ahler and hyper-K\"ahler geometries are characterised in terms
of properties of the K\"ahler forms.  We construct twistor spaces associated with HKT
spaces and state a reconstruction theorem, thus generalising the twistor construction
and the reconstruction theorem of ref. [\sal,\roc] for hyper-K\"ahler manifolds.  Finally we 
use the above results to give a geometric interpretation of the (4,0) superfields introduced in [\hpa] as
holomorphic bundle maps from (4,0) harmonic superspace [\dks] into the twistor space of the sigma
model target manifold and we exploit the complex structure of the twistor space to
construct a (4,0) action in terms of these superfields.

A complex  manifold $M$, with metric $g$, complex structure $I$ and a three form $H$,
 has a KT structure provided that these tensors obey the following
conditions:
$$\eqalign{
I_i{}^k I_j{}^{l} g_{kl} &=g_{ij}
\cr
\nabla^{(+)}_i I_j{}^k&=0\, ,}
\eqn\inaone
$$
where the connections, $\Gamma^{(\pm)}$, of the covariant derivatives, $\nabla^{(\pm)}$, are given by
$$
\Gamma^{(\pm)}_{jk}{}^i=\Gamma_{jk}{}^i\pm{1\over2}H_{jk}{}^i\, ;
\eqn\inatwo
$$
$\Gamma$ is the Levi-Civita connection of the metric $g$ and 
$$
H\equiv{1\over3}dx^k\wedge dx^j\wedge dx^i H_{ijk}
\eqn\inathree
$$
is a three-form on the manifold $M$.  If  no further conditions are imposed on $H$, we say that
the manifold $M$ with tensors $g,I$ and  $H$ that satisfy \inaone\ has a weak KT structure.
However, if in addition we take $H$ to be a closed three form ($dH=0$), we say that $M$ has a
strong KT structure, in which case we can write
$$
H={1\over 3} db
\eqn\inathree
$$
for some locally defined two-form $b$ on $M$.\foot {We use superspace form notation where the
exterior derivative, $d$, acts from the right.}  Finally, if $H$ is the zero three-form, the manifold
$M$ becomes K\"ahler.
The target space, $M$, of a (2,0)-supersymmetric sigma
model with torsion is a manifold with a strong KT structure. The couplings of the classical action of the
theory are the metric,
$g$, of $M$ together with the two-form $b$.  However, in the quantum
theory  and in particular in the context of the anomaly cancellation
mechanism [\hw,\sen, \hpc], the (classical) torsion $H$, \inathree, of
(2,0)-supersymmetric sigma models receives corrections proposional to
the Chern-Simons three-form of the  $\Gamma^{(-)}$ connection. 
Therefore the new torsion is not a closed three form but rather
$dH=c\,{\rm tr} R^{(-)}\wedge R^{(-)}$ for some constant, $c$; we have
used the same notation for the torsion before and after the
redefinition.  Therefore, although classically the the target space of
(2,0)-supersymmetric sigma models has a strong KT structure, quantum
mechanically this changes to a weak KT structure, albeit of a
particular type.

A Riemannian manifold, $M$, with metric $g$ and torsion a three-form, $H$, has an HKT structure if it
admits three (integrable) complex structures $\{I_r; r=1,2,3.\}$, that obey the following conditions
$$
\eqalign{
I_r I_s&=-\delta_{rs}+\epsilon_{rst} I_t
\cr
I_{r i}{}^k I_{r j}{}^{l} g_{kl} &=g_{ij}\ ; \qquad r=1,2,3
\cr
\nabla^{(+)}_i I_{rj}{}^k&=0\ ,}
\eqn\inone
$$
where the connections, $\Gamma^{(\pm)}$, are given in  \inatwo.  It is evident that if $H=0$, then the
conditions \inone\ are those of hyper-K\"ahler geometry. As in the case of KT structures, we can define a strong HKT structure and a weak HKT structure depending on
whether or not the three-form $H$ is closed. Both strong and weak HKT geometries arise in the context
of (4,0)-supersymmetric sigma models with torsion.  The strong HKT geometry is the geometry of the sigma
model manifold in the classical theory, while the weak geometry is the geometry of the sigma model
manifold in the quantum theory as explained for the case of the (2,0)-supersymmetric sigma model
above.
There are many examples of manifolds with strong HKT structures. These include group
manifolds [\vp, \iv]
with $SU(2)\times U(1)$ as the simplest example. One can construct other four-dimensional examples
by starting from hyper-K\"ahler manifolds with metric $g_h$ and then setting $g=e^F
g_h$ and 
$H={}^*dF$.  The metric
$g$ and torsion $H$ describe a strong HKT structure provided that $e^F$ is a harmonic function with
respect to the metric $g_h$ [\chs]. Furthermore, the conditions \inone\ can be
solved exactly if one assumes that the four-manifold $M$ admits a triholomorphic Killing vector field
which in addition leaves the torsion $H$ invariant [\gp].   The associated strong HKT geometry is naturally
associated with monopoles  on the round three-sphere and an example of such geometry is the Taub-NUT
geometry with non-zero torsion found in  refs. [\dv, \bv]. In the limit that the torsion
vanishes, the strong HKT geometry of [\gp] becomes that the Gibbons-Hawking
hyper-K\"ahler geometry [\gh, \gwg]. The Gibbons-Hawking metrics are associated with monopoles on the
Euclidean three-space.  
The conditions \inaone\ and \inone\ on the various tensors associated with manifolds with a KT and
HKT structure, respectively,  can be rewritten in terms of exterior differential relations.  This is
the analogue of a similar situation that arises in the case of K\"ahler and hyper-K\"ahler manifolds
where the covariant constancy condition of a complex structure is equivalent to the symplectic
condition for the associated K\"ahler form.  However due the the presence of torsion, the exterior
differential relations for KT and HKT manifolds are somewhat different from those of K\"ahler and
hyper-K\"ahler manifolds. 

We first consider the exterior differential relations for weak KT manifolds. For this,  we use 
notation similar to that of ref. [\hpb] and introduce the inner derivation,  $\i_I$, and the exterior
derivation, $d_I$, associated with the complex structure $I$ as follows:
$$\eqalign{
\i_I \pi&=p dx^{i_p\dots i_1} I_{i_1}{}^j \pi_{ji_2\dots i_p}
\cr
d_I\equiv d'&=\i_I d-d\i_I
\, ,}
\eqn\inaaone
$$
where
$$
\pi=dx^{i_p\dots i_1}\pi_{i_1i_2\dots i_p}
\eqn\inaatwo
$$
is a p-form.
Using the first equation in \inaone,  we introduce a two-form $\omega$ as follows:
$$
\omega(X,Y)=g(X,YI)\ .
\eqn\inaathree
$$
Then the covariant constancy condition in \inaone\ and the fact that $H$ is a (2,1) and (1,2) form
with respect to $I$ implies that
$$
H=d'\omega\ .
\eqn\inaafour
$$
The above statement has a converse:  if $M$ is a complex manifold with complex
structure $I$ and a non-degenerate two-form $\omega$ which is hermitian with respect to $I$, then
$M$ admits a weak KT structure with metric, $g$, given in \inaathree\ and torsion, $H$, given in
\inaafour.  To show this, one makes use of the vanishing of the Nijenhuis tensor of the complex structure $I$.
To describe the geometry of manifolds with a strong KT structure in terms of the exterior differential
relations one has to impose, in addition, the constraint that $H$ be closed which implies that $\omega$ should satisfy:
$$
dd'\omega=0\ .
\eqn\inaafive
$$
We remark for use later in the letter that for manifolds with a strong KT structure the metric $g$ and the
locally defined two-form $b$ can be expressed in terms of a (real) one-form potential $k$, [\hw].  If we introduce complex co-ordinates $\{z^\alpha; \alpha=1,\dots,n\}$ (${\rm dim }M=2n$) on $M$
with respect to the complex structure $I$ we can write  
$$\eqalign{
g_{\alpha\bar\beta}&=\partial_\alpha k_{\bar\beta}+\partial_{\bar\beta} k_\alpha
\cr
b_{\alpha\bar\beta}&=\partial_\alpha k_{\bar\beta}-\partial_{\bar\beta} k_\alpha \ .
}
\eqn\kone
$$
Next we consider the case of manifolds with a weak HKT
structure. We first introduce three inner derivations,
$\i_r$, and three exterior derivations,
$d_r$, associated with the three complex structures, $I_r$.   These derivations together with the
exterior derivative
$d$ satisfy the differential  algebra
$$
\eqalign{
\i_r\i_s-\i_s\i_r&=2\epsilon_{rst} \i_t \cr
\iota_r d-d\iota_r&=d_r
\cr
\iota_r d_s- d_s\iota_r&=-\delta_{rs} d+\epsilon_{rst} d_t\ , 
\cr
d^2&=0,\cr
d_rd_s+d_s d_r&=0
\cr
dd_r+d_rd&=0\ .}
\eqn\drone
$$
To derive the differential algebra \drone, we have used the algebraic relations \inone\ and the
integrability properies of the complex structures $\{I_r; r=1,2,3\}$.
We then introduce the three two-forms, $\{\omega_r\, ; r=1,2,3\}$,   as in \inaathree , one for
each of the three complex structures $\{I_r;\, r=1,2,3\}$.
Using the covariantly constancy condition in \inone\ and the integrability conditions of the
complex structures, we can show that $H$ is the sum of a three-form of type (2,1)  with respect to all complex structures and its complex conjugate which is of type (1,2).
structures.  This fact can be summarised as follows:
$$
\i_r\i_sH=-\delta_{rs} H+\epsilon_{rst} \i_tH\ .
\eqn\drtwo
$$
Using this equation, the differental algebra \drone\ and the covariant constancy condition in
\inone,  we can show that 
$$
d_r\omega_s=\delta_{rs}H-\epsilon_{rst} d\omega_t\ .
\eqn\drthree
$$
Observe that the diagonal conditions ($r=s$) in the above equation imply the off-diagonal ones
($r\not=s$) and vice versa.  This will be used later in the twistor construction for HKT manifolds.
For manifolds with a weak HKT structure the above has a converse that can be stated as
follows: let $M$ be a manifold with $\{I_r;r=1,2,3\}$ complex structures that obey the algebra of
imaginary unit quaternions and suppose that there exists a non-degenerate two-form $\omega_3$ which is (1,1) with respect to $I_3$ and which satisfies 
$$
\i_1\i_2\om_3 =0\ ,
\eqn\drthreea
$$
then three two-forms, $\om_r$, can be defined which satisfy the relations
$$
\i_r\om_s=2\epsilon_{rst}\om_t\ ,
\eqn\drthreeb
$$
and from any one of which one can construct the trihermitian metric $g$ by
$$
\omega_r(X,Y)=g(X,YI_r).
\eqn\drfive
$$
$M$ admits a weak HKT structure provided that, in addition, the 3-form $H$
defined by $H=d_3\om_3$ is (2,1) plus (1,2) with respect to all complex structures. To describe the differential relations for manifolds with a strong HKT structure,
we should also impose the condition:
$$
dd_r\omega_r=0 \, , \qquad r=1,2,3\ ,
\eqn\drsix
$$
for the torsion $H$ to be a closed three-form on $M$.
To construct the twistor space of HKT manifolds, we first observe that on any manifold, $M$, with
three complex structures,
$\{I_r; r=1,2,3\}$, that satisfy the algebra of imaginary unit quaternions, the tensor
$$
{\I}=a_r I_r; \qquad \qquad a_r a_r=1
\eqn\tone
$$
is also a complex structure.  Thus there is an $S^2$'s worth of complex structures on $M$, and
the twistor space is simply $Z=M\times S^2$.  If $(x,y)\in Z$ where $y$ denotes the usual (affine)
complex co-ordinate on $S^2=\cp^1$.  Then we have
$$
T_{(x,y)}Z=T_xM\oplus T_yS^2
\eqn\ttwo
$$
and so we can define an almost complex structure on $Z$ by
$$
\hat{\I}=({\I}, I_0)
\eqn\tthree
$$
where $I_0$ is the complex structure on $\cp^1$ and
$$
{\I}={1\over 1+y\bar y}\big[ (1-y\bar y) I_3+(y+\bar y) I_1+i (y-\bar y) I_2\big]\ .
\eqn\tfour
$$
In fact $\hat{\I}$ is a complex structure.  To see this let $\phi$ be a (1,0) form on $M$
with respect to $I_3$, $I_3\phi=i\phi$, then 
$$
\hat\phi = (1-iy\i_1)\phi
\eqn\tfive
$$ 
is (1,0) with respect to ${\I}(y)$ as is not difficult to show.  Now $\hat{\I}$ is
integrable if the exterior derivative (on $Z$) of any form $\hat \phi$ which is (1,0) with respect to
$\hat{\I}$ is the sum of terms each of which is the wedge product of an arbitrary one-form
with a (1,0) form, i.e. if
$$
d\hat \phi=\sum_p \lambda_p\wedge \rho_p
\eqn\tsix
$$
on $Z$, where each $\rho_p$ is (1,0). Clearly $dy$ is (1,0) with respect to $\hat{\I}$ and satisfies \tsix\ so we only need to
check \tsix\ for (1,0) forms of the type \tfive\ now interpreted as forms on $Z$.  It is not
hard to show that
$$
d\hat \phi=idy\wedge \iota_1\phi + {1\over2}dx^j\wedge dx^i H_{ij}{}^k\hat\phi_k + dx^j\wedge
dx^i
\nabla^{(+)}_{i} \hat\phi_j
\eqn\tseven
$$
The first term on the RHS of \tseven\ is obviously of the desired form as
is the second, due to the fact that $H$ is (2,1) plus (1,2) and $\hat\phi$ (1,0) with respect to
$\hat{\I}$. Finally, it is easy to check that the third term has no (0,2) part either due to the fact that $\I$ is covariantly constant with respect to $\nabla^{(+)}$. Hence $Z$ is complex. 

Having constucted the twistor space $Z$ of a manifold $M$ with an HKT structure, we shall now reverse the procedure and
determine the metric and torsion of $M$ from data on the twistor space. As we have
shown, $Z$ is a complex manifold and so we can write $TZ\otimes \C=\tau\oplus\bar\tau$, where $\tau$
is the holomorphic tangent bundle. Since the projection $p: Z\rightarrow \cp^1$ is
holomorphic, we define $\tau_f={\rm Ker}\ dp|_{\tau}$. The holomorphic sections of the bundle
$Z\rightarrow \cp^1$ are the twistor lines and the manifold $M$ can be thought as the space of
their deformations (the space of twistor parameters).  The normal bundle of every twistor line is
isomorphic to $\C^{2n}\otimes O(1)$ where $O(1)$ denotes the twist of the normal bundle over
$\cp^1$. (The $O(1)$ twist of the normal bundle of the twistor line is related to the fact that
the form \tfive\ is linear in $y$.)   One then can define  an (2,0)-form
$\omega$ as follows:
$$
\omega=-(\omega_1-i\omega_2)+2y \omega_3+y^2 (\omega_1+i\omega_2)\ .
\eqn\tnine
$$
This form is a section, holomorphic with respect to $\cp^1$, of the bundle 
$\Lambda^2 \tau^*_f(2)$, where the number $2$ denotes the twist of the bundle over
$\cp^1$ and it is related to the fact that $\omega$ is quadratic in the $y$
co-ordinate.   Note also that there is the real structure, $r:M\times \cp^1\rightarrow
M\times \cp^1$, on $Z$ defined as follows:
$$
r:(x,y)\rightarrow (x,-{1\over \bar y})\ .
\eqn\tten
$$
The twistor lines and the form \tnine\  are compactible with the real structure $r$. 
 Moreover the
real structure $r$ transforms the complex structure ${\I}$ to $-{\I}$.   
Now we are ready to state the reconstruction theorem for manifolds with a weak HKT 
structure.  This is
as follows:  let $Z$ be a complex manifold with complex dimension
$2n+1$ and the following properties: 1. Z is holomorphic fibre bundle $p: Z\rightarrow \cp^1$,
2. the bundle admits a family of holomorphic sections each with normal bundle isomorphic to
$\C^{2n}\otimes O(1)$, 3. there is a section $\omega$, holomorphic with respect to $\cp^1$,  of
$\Lambda^2\tau^*_f$ defining a non-degenerate two-form at each fibre that satisfies
$$
(id+d_{\I})\omega=0 \ ,
\eqn\televen
$$
4. $Z$ has a real structure $r$ compactible with the above data and inducing the antipodal map
on $\cp^1$.  Then the parameter space of real sections is a 4n-manifold, $M$, with a natural weak HKT
structure.  
Many steps in the proof of the above theorem are similar to those of the reconstruction theorem
for hyper-K\"ahler manifolds [\roc].   The main difference is the condition, \televen,
that the two-form, $\omega$, satisfies \foot {In the hyper-K\"ahler case the
condition on $\omega$ is $d\omega=0$.}.  To derive the HKT
structure on the space of parameters of the twistor lines from  \televen, one 
evaluates \televen\ at the points $\{1,-1,i,-i,0, \infty\}$ of $\cp^1$
and
 then 
observes that the
resulting conditions imply the off-diagonal, ($r\not=s$), conditions of \drthree. 
The metric,
$g$, on $M$ is defined as in the hyper-K\"ahler case and the torsion is defined as 
follows:
$$
H=d_1\omega_1\ .
\eqn\ttwelve
$$
Finally, using the equivalence of the diagonal and the off-diagonal conditions of 
\drthree\ and the
relation of the exterior differential relations \drthree\ to the weak HKT structures,
 one sees that the space of parameters, $M$, of the real twistor lines has a weak HKT
structure.  We can also incorporate strong HKT structures.  The only difference
between the reconstruction theorems for manifolds with weak and strong HKT structures
 is the condition on the form $\omega$.  In the strong case one should require, in
addition to \televen\ ,  that
$$
(id+d_{\I})\partial\omega=0\ ,
\eqn\tthirteen
$$ 
where $\partial$ is the exterior derivative along the $y$ direction. This condition is what is
needed to show that the torsion \ttwelve\ is a closed three-form on the space of twistor
parameters, $M$.

Now consider a (4,0) supersymmetric sigma model in (4,0) superspace $\Sigma$.
This space has coordinates $(u,v,\th_o,\th_r)$, where $(u,v)$ are light-cone
cordinates for two-dimensional Minkowski space, and where the
supercovariant  derivatives satisfy 
$$
\eqalign{
[D_o,D_o]&= i\partial_u \cr
[D_o,D_r]&= 0 \cr
[D_r,D_s]&= i\delta_{rs}\partial_u\ . }
\eqn\tfourteen
$$
The sigma model superfield is a map from $\Sigma$ to $M$ which satisfies
$$
D_r X^i=-D_o X^j I_{r j}{}^i
\eqn\tfifteen
$$
as a consequence of which the action
$$
A=-2i \int\, du dv D_o\, \{(g+b)_{ij} D_o X^i \partial_v X^j\}
\eqn\tsixteen
$$ 
is (4,0)-supersymmetric [\hpa]. 
The above superspace is not a complex space but it does admit several $CR$-structures [\rs,\hh],  which
can be thought of as partial complex structures. More precisely, a real 
(super)manifold of
dimension $2n+m$, where $m,n\in \Z$ (or
$\Z^2$ in the super case), is a $CR$ (super)manifold if the complexified tangent 
bundle has a
complex rank $n$ sub-bundle which is involutive. In other words, there must be $n$ 
(local) linearly independent complex vector fields which form a closed system under
 Lie
brackets. There are many odd $CR$ structures on $\Sigma$, i.e. $CR$ structures
generated by odd vector fields, and they can be understood in terms of complex
structures of the odd tangent bundle. In fact, (4,0) superspace has  a natural set of
three fibre complex structures ${J_r}, r=1,2,3,$ obeying the algebra of the unit
imaginary quaternions. With the above covariant derivatives as a basis of odd tangent
vectors, the components of the $J$'s can be taken to be
$$
\eqalign{
(J_r)_{0s}&=-\delta_{rs} \cr
(J_r)_{st}&=-\epsilon_{rst} \cr}
\eqn\tseventeen
$$
with the remaining components being determined by antisymmetry, since the standard 
Euclidean metric is trihermitian. The $CR$ derivative associated with any of these
complex structures has components given by ${1\over2}(1+i J_r)D$, where $D$ denotes
the set of covariant derivatives. The algebra of the $J$'s and the algebra of the
$D$'s then ensures that these derivatives do indeed anticommute amongst themselves.
Clearly, $\J:=a_r J_r$, where $a_r a_r=1$, is also an odd complex structure,
 so that there is an $S^2$ of such $CR$ structures on $\Sigma$, and hence we can form
the twistor space, $\hat\Sigma=\Sigma\times S^2$, in an analogous fashion to the
twistor space associated with $M$. This is in fact the (4,0) harmonic superspace discussed in from a different perspective in [\dks].
It is not difficult to show that $\hat\Sigma$ is a $CR$ supermanifold with $CR$
structure of rank $(1|2)$; the corresponding $CR$ derivatives are those given above
(for the structure $\J$), together with ${\partial\over\partial\bar y}$, where $y$ is
the standard holomorphic coordinate on $S^2$. The twistor space can be considered as a
fibre bundle over $\cp^1$ where the fibre at $y$ is $\Sigma$ together with the
$CR$-structure determined by $\J(y)$. This space is not trivial as a $CR$ bundle as
the (two-dimensional) complex odd part of the fibre has twist 1 with respect to
$\cp^1$. There are two independent complex components of any $CR$ derivative; using
the above prescription for computing them one finds that, for $\J$, one of them is
$$
\bar D(y):= D_o-ia_r D_r\ .
\eqn\teighteen
$$
This derivative does not commute with ${\partial\over\partial\bar y}$, but instead 
the commutator gives a new odd vector field, $\bar D'(y)$, and these three vector
fields form a basis of the $CR$ structure. It therefore follows that any function $f$
on $\hat \Sigma$ which is analytic with respect to $\cp^1$ and which satisifies
$$
\bar D(y) f=0\
\eqn\tnineteen
$$
is in fact $CR$-analytic. These are precisely the type of fields we are interested 
in because the sigma model constraint can be rewritten as
$$
a_r D_r X^i =-D_o X^j \I_j{}^i.
\eqn\ttwenty
$$
In complex coordinates $Z^{\alpha}$ with respect to $\I$ this is just
$$
\bar D(y) Z^{\alpha}=0
\eqn\ttwentyone
$$
Since the coordinates $Z^{\alpha}$ do not depend on $\bar y$, it follows that
the sigma model map is a $CR$-analytic map from $\hat\Sigma$ to $Z$, which is in
 addition 
fibre-preserving and which induces the identity on the base space, $\cp^1$. Note that the superfields contructed here are short multiplets, in contrast to those of ref. [\dks] which are not analytic with respect to $\cp^1$. 

This construction allows us to write a new form of the (4,0) action. In the (2,0) 
case, one has the same action but with the fields now being (2,0) superfields
satisfying the constraint
$$
D_1 X^i=-D_o X^j I_j{}^i
\eqn\ttwentytwo
$$
Switching to complex coordinates, using the above constraint and the expression 
for the Kahler
form in terms of the potential $k$ given in \kone\ one arrives at the manifestly
(2,0) invariant
form of the action
$$
A=-i\int du dv D\bar D\{ k_{\alpha} \partial_v Z^{\alpha}-
\bar{k}_{\bar\alpha} \partial_v Z^{\bar\alpha}\}\ ,
\eqn\ttwentythree
$$
where $D=D_o+iD_1$.
We can carry out exactly the same construction in the (4,0) case using the two-form 
$\Omega:=a_r \om_r$. That is to say, for each point $y\in \cp^1$, we 
have a (2,0)
 sigma model
with (2,0) derivative $D(y)$ and potential $k(y)$, and the action can 
therefore be
 converted
into the form
$$
A=-i\int du dv D(y)\bar D(y)\{ k_{\alpha}(y) \partial_v Z^{\alpha}-
\bar{k}_{\bar\alpha}(y)
\partial_v Z^{\bar\alpha}\}\ ,
\eqn\ttwentyfour
$$
which appears at first sight to depend on $y$, although it clearly cannot by 
construction.

\vskip 1cm
\noindent{\bf Acknowledgments:}  G.P. is supported by a University Research
Fellowship from the Royal Society.
\refout
\end